\documentstyle[aps,multicol,amstex,natbib]{revtex}
\input epsf
\def\etal{\textit{et al.}}
\begin{document}
\title{Competing Populations in Flows with Chaotic Mixing}

\author{Istv\'an Scheuring$^\ddag$,Gy\"orgy K\'arolyi$^\star$,
Zolt\'an Toroczkai$^\P$,
Tam\'as T\'el$^\S$,
and \'Aron P\'entek$^\dag$}

\address{
$^\ddag$Department of Plant Taxonomy and Ecology,
Research Group of Ecology and Theoretical Biology,
E\"otv\"os University,
Ludovika t\'er 2, H-1083 Budapest, Hungary\\
$^\star$Department of Structural Mechanics,
Budapest University of Technology and Economics, M\H{u}egyetem rkp.~3,
H-1521 Budapest, Hungary\\
$^\P$Theoretical  Division and Center for Nonlinear Studies,
Los Alamos National Laboratory,
Los Alamos,  New Mexico 87545, USA \\
$^\S$Institute for Theoretical Physics,
 E\"{o}tv\"{o}s
University, P. O. Box 32, H-1518 Budapest, Hungary\\
$^\dag$Marine Physical Laboratory, Scripps Institution
of Oceanography, University of California
at San Diego, La Jolla, CA 92093-0238, USA}

\date{\today}
\maketitle
\begin{abstract}
We investigate the effects of spatial heterogeneity on the
coexistence of competing species in the
case when the heterogeneity is dynamically generated
by environmental flows with chaotic mixing properties.
We show that one of the effects of chaotic advection on the
passively advected species (such as phytoplankton, or
self-replicating macromolecules) is the possibility of coexistence
of more species than that limited by the number of niches they occupy.
We derive a novel set of dynamical equations
for competing populations.
\end{abstract}

\vspace*{0.6cm}

\begin{multicols}{2}

\section*{Introduction}

One of the classical problems of ecology is the identification of the
mechanisms responsible for the coexistence of competing species.
It is an observational fact that in Nature numerous
species are able to coexist, all competing for a limited number
of resources. This observed coexistence is at odds with many
classical theories and empirical studies predicting
competitive exclusion of all but the
most perfectly adapted species for each limiting factor
\citep{Gau35,Har60}. However, one of the
key ingredients in these classical studies was the
assumption of a homogeneous, well mixed and non-structured
environment which leads to an equilibrium state in the system.
Thus, if coexistence is to persist over longer time periods,
it must have a nonequilibrium character.

The coexistence problem is best illustrated in the case
of \emph{phytoplankton communities} as
was originally
presented by \citet{Hut61}.
Here a number of species coexist
in a relatively isotropic or unstructured environment, all
competing for the same sorts of materials, and the number of species exceeds
considerably the number of limiting factors. 
To solve this so called ``paradox of plankton'', Hutchinson 
put forward the 
idea that seasonal environmental 
changes prevent competitive 
exclusion in natural phytoplankton communities. 
Thus the species of the community, at least on 
the time scale of ecological observation, are 
in nonequilibrium coexistence. 

Since then numerous investigations revealed  
many different 
mechanisms, including spatial and 
temporal heterogeneity of habitat,
predation, disturbance, coevolution, etc.\ \citep{Wil90,Che00},
increasing
 the probability of competitive coexistence.
Naturally, under the word ``competition''  many different
biological phenomena are collected together, which influence
the coexistence of species in different ways.

Thus the original problem changed into finding the most relevant
mechanisms which maintain diversity
in particular situations \citep{Con78,Hus79,Wil90,Til93,Bar97}.
Despite the vivid debate in this field
of ecology, there is  by now a consensus
that climatic periodicities and fluctuations
play the main role in causing
species' persistence in phytoplankton
communities \citep{Gae86,Rey93,Som93}.
It is frequently argued that an intermediate disturbance
\citep{Con78} is the most adequate hypothesis for the
explanation of high diversity in aquatic
systems \citep[cf.][]{Rey98}.

One can meet a similar problem in \emph{early evolution of life}.
Since life evolves from the simple structured entities
to the most complex ones, there must have been
a stage in the evolution, when life was essentially no
more complex than what a collection  of self-replicating
nucleic acids present \citep{May95}.
They were competing for a few limiting resources
(such as mononucleids and
energy rich chemicals) and making copies of
themselves without any specific
enzyme. Without enzymes the copying accuracy could not be very high.
Estimating the selective superiority
of the best replicator and the copying accuracy per
nucleotide, it is concluded  that the maximum length of these
molecules is about 100 nucleotides \citep{Eig71}.
However, in a well mixed homogeneous environment, as the
prebiotic ocean  is supposed to have been,
there are only a few  winners of the
selection, namely the most fit macromolecule surrounded
by its closest mutants
\citep{Eig71,Eig79}.
But how can we surmount the gap  between
these primitive replicators with 100 nucleotides
and the most simple
RNA viruses with 4000--5000 nucleotides?
Specific replicase enzymes are needed to increase
the copying fidelity, and thus the length of the
replicator, but these replicators are too short to code
specific enzymes. This is the ``Catch~22'' of the
prebiotic evolution \citep{May83}: no genome without
an enzyme, however no enzyme without genomes.
This problem can be resolved if some mechanism maintains
the coexistence of several  different replicator molecules
and therefore the information necessary for coding a
replicase enzyme can be stored by the union of smaller
information carriers. In this situation the replication error
does not grow exponentially as in the case of a base-by-base
copying, it grows only linearly with the number of
smaller carriers.

\begin{figure}[htbp]
\vspace*{-0.5cm}
\hspace*{0cm}\epsfxsize = 3.2 in \epsfbox{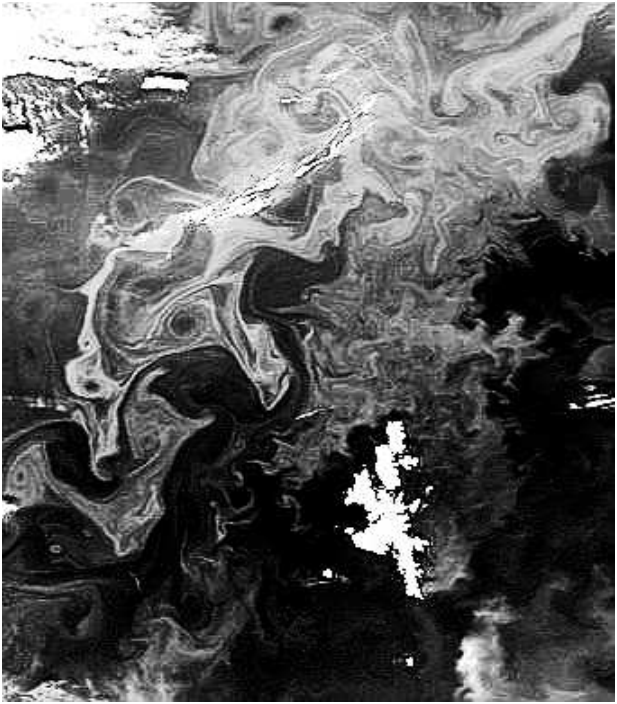}
\vspace*{0.5cm}
\caption{ SEAWIFS image of a phytoplankton bloom at
Shetland Islands, May 12, 2000,
from the NASA archive. Plankton individuals (light grey)
move along a fractal set. }
\label{fig:satelite}
\end{figure}

Current theories point out coexistence of replicators moving on a
surface 
\citep{Boe91,Cza00}, preferring thus the concept of
``prebiotic pizza'' against
the concept of ``prebiotic soup'' \citep{Wac94}. However some
kind of cooperation among the replicator molecules are
built into these models,
consequently they are not completely competitive.
An alternative explanation assumes that both the
replicative and enzymatic
functions were co-evolved, thus the lenght of the
replicators and the accuracy
of enzymatic functions increased together
 \citep{Poo99,Sch00a}.
In both problems (i.e., in the paradox of plankton
and in the Catch~22
of prebiotic evolution) the  traditional population
dynamical equation
for two species $B_1$, $B_2$ competing for the
resource $A$ read:
\begin{eqnarray}
&&\frac{dN_1}{dt} = \alpha_1 N_1 - \delta_1 N_1, \\\label{trad0}
&&\frac{dN_2}{dt} = \alpha_2 N_2 - \delta_2 N_2.\label{trad}
\end{eqnarray}
Here $N_i$ is the instantaneous
number of individuals of species $B_i$ in a given range
of a well strirred region. The instantaneous
parameters $\alpha_i$, $\delta_i$
are positive and
depend, in general,  on the concentration of the resource material
$A$, too.

\begin{figure}[htbp]
\begin{minipage}{3.3in}
\hspace*{-0.4cm}\epsfxsize = 3.3 in \epsfbox{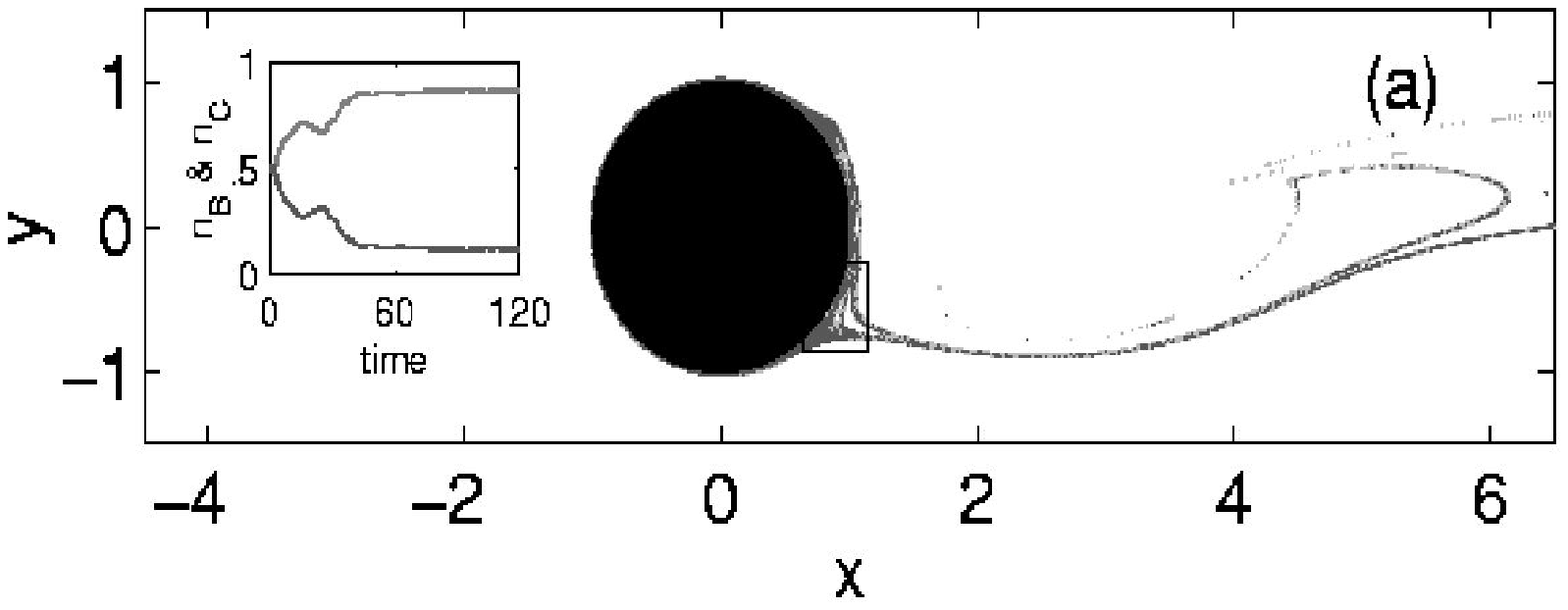}
\end{minipage} \\
\begin{minipage}{3.3in}
\hspace*{-0.4cm} \epsfxsize = 3.3 in \epsfbox{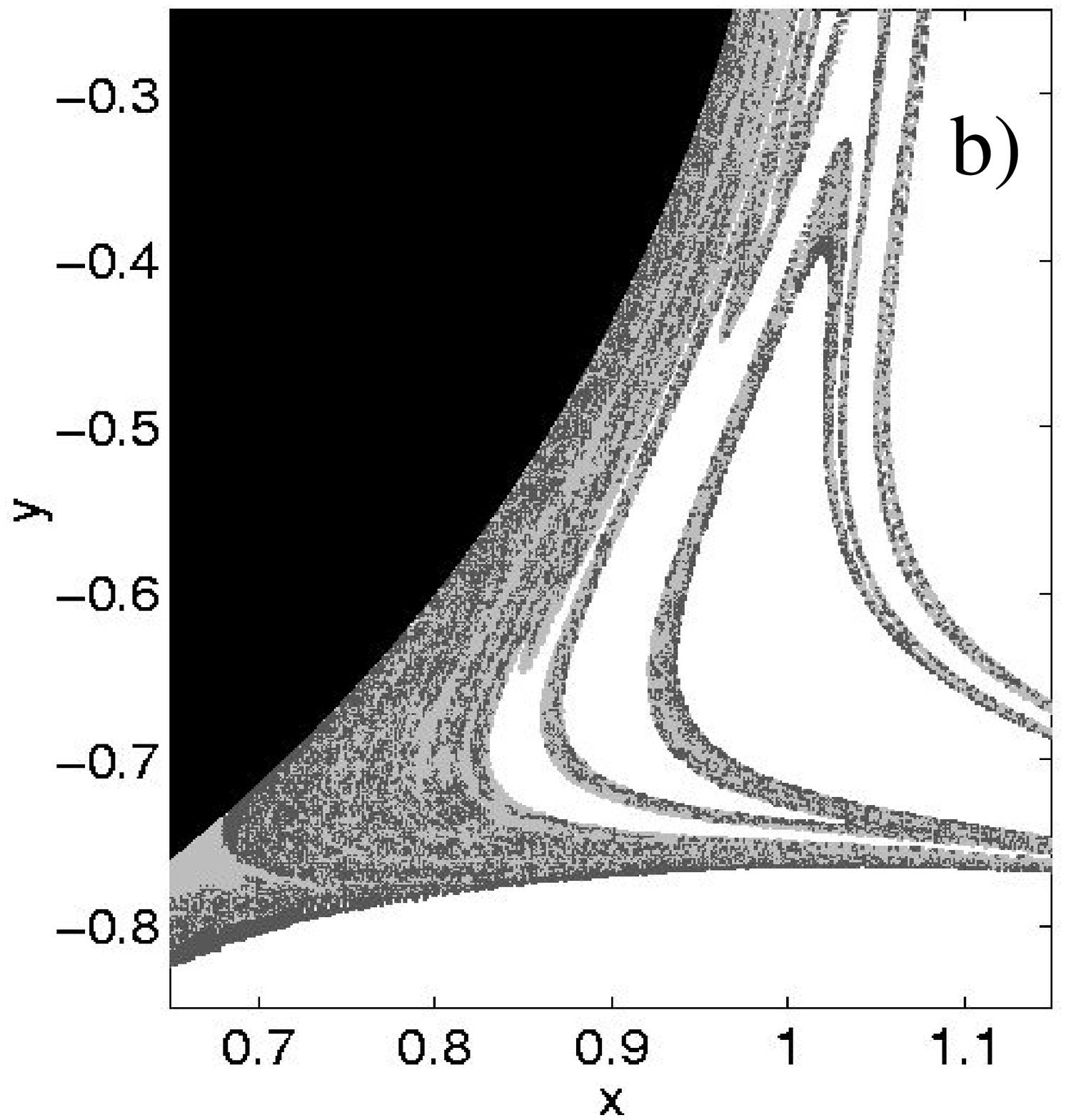}
\end{minipage}
\vspace*{0.0cm}
\caption{The distribution of two populations (light gray
($B$) and dark gray ($C$))
competing for the same resource material (white ($A$))
in the wake of
a cylinder. The flow is from left to right. The inset
in (a) shows the
time-dependence of the population numbers $n_B,n_C$ and
clearly indicates
the approach to a steady state of coexistence after about
$40$ time units which is
the period of the flow. A blowup of the region
indicated by a rectangle
in (a) is seen in (b).  Species distribution is
strikingly similar to many of the patterns
found on the NASA SEAWIFS satelite pictures of plankton blooms
(see Fig \ref{fig:satelite}).
After  \citet{Kar00}.}\label{fig:Compare}
\end{figure}

Independently of the particular form of this
dependence and the dynamical equation of $A$,
\emph{no} fixed points can exist in the system
in which both species would be in a steady state with
nonzero values of $N_i=N_i^*$ \citep{Gur98}.
The coexistence might, however,
be possible in \emph{imperfectly stirred environments}. 
As numerous remote sensing images demonstrate
phytoplankton are distributed along fractal
filaments in the oceans indicating strong but
imperfect mixing environment
(Fig \ref{fig:satelite}).

Recent development in the field of hydrodynamics
encouraged us
to revisit the population dynamics of competing species in \emph{open} 
aquatic
systems. In aquatic systems of large extension, on
the time scales characteristic to the life cycle
of microorganisms and replicators, the hydrodynamical flows are
locally open, i.e. there is a net current, transporting both
competitors and nutrients,
flowing through the typical observation region.
It is even more obvious that the flow
is open in the wake of islands surrounded by strong
ocean currents \citep{Ari97} and around the deep see hot springs
where the cradle of life probably swung \citep{Hol92}.

With the aid of numerical simulations we have
previously shown  that the
coexistence of passively
advected competing species is typical in open chaotic flows 
\citep{Kar00,Sch00b}. For simplicity, we have considered the
two-dimensional
flow around a cylindrical obstacle
placed into a uniform background flow.
For moderate inflow velocities
there is a periodic detachment of vortices in the wake of the obstacle
with period $T$, which forms the von K\'arm\'an
vortex street \citep{Sha91,Jun92,Som96}. 
The flow in the wake is time-dependent but still spatially
regular. Here individuals of two passively advected species compete for a
common limiting resource, see Fig.~\ref{fig:Compare}.
We argued that
coexistence is due to the fractal structures typically appearing in the
advection patterns of such flows, however, we have given  only a  heuristic
interpretation for the mechanisms maintaining coexistence in this
hydrodynamical system. In this article we present a  mathematical deduction to
explain coexistence of competitors in open chaotic flows.   The mathematical
problem is to investigate two coupled population dynamical processes
evolving on a \emph{fractal} support.
We shall present a new class of equations which describe this situation,
and allow for the coexistence of at least two species
competing for the same resource. A novel feature of these equations  
will be a singularly nonlinear 
(power-law) form of both the replication term and the 
coupling between the populations.

In the following section we summarize the
qualitative features of the
relevant physical process, followed by a  study of the
dynamics of a single population in an open chaotic flow. Consequently, the
coexistence of competitors is discussed by first giving
a qualitative argument based on the single population
picture, followed by a detailed mathematical model
leading to the aforementioned new type of population dynamical equations.
Next, this theory is compared with further numerical results carried out
on a simple map modelling the advection dynamics, on the so-called baker map.
 We conclude with a summary and outlook.

\section*{Passive advection in open flows}

Chaotic advection in open hydrodynamical flows is an 
ubiquitous phenomenon.  A flow is considered locally open 
if there is a  net current flowing through 
the observation region \citep{Lam32}. 
It became clear in the last decade 
that passive advection even in simple 
time-dependent flows is typically 
chaotic \citep{Pen96,Som96,Kar97} and possesses  complicated 
particle trajectories.
These flows, characterized by strong \emph{imperfect} mixing, 
lead to a fractal spatial distribution of
advected particles in a finite region of the flow.
This region is called the \emph{mixing region}.
In our terminology, a flow is chaotic if the advection dynamics
generated by the flow is chaotic.

In the case of several (three or more) types of passively advected tracers,
distinguished for example by their color, it was shown
\citep[and references therein]{Tor97} that
their distribution may follow a rather non-trivial topology on the
fractal, a property called the Wada property: every point \emph{on}
the fractal is lying on the boundary of at least three colors.

It is worth emphasizing that a complicated flow field (turbulence)
inside the mixing region
is not required for the flow to be chaotic, (i.e., for
complex advection dynamics or for the appearance of
fractal patterns).
Even simple forms of time dependence,
e.g.\ a periodic repetition
of the velocity field with some period $T$, is sufficient \citep{Are94}.
Thus, for sake of simplicity, 
we examine advection in time-periodic open flows.

The complicated form of trajectories implies a long time spent
in the mixing region. In other words, advected particles can be temporarily
trapped there.
It is even more surprising, however,  that
there is an \emph{infinity} of special
nonescaping orbits.
The simplest among these orbits are the periodic
ones with periods that are integer multiples of the
flow's period $T$.
All the nonescaping orbits are highly unstable, of saddle
type, and possess a
strictly positive local Lyapunov exponent
(which is the expanding eigenvalue of the unstable periodic orbit).
Another important feature of these orbits is that
despite their infinite number
they are rather exceptional so that they cannot fill a finite portion
of the phase space. Indeed,
the union of all nonescaping orbits forms
a fractal ``cloud'' of points on any snapshot.
This fractal cloud moves
periodically with the flow and never leaves the mixing region.

Typical advected particle trajectories are not in the set of
the nonescaping orbits, but are, nevertheless, influenced by them.
They follow closely some periodic orbit
for a while and later turn to follow others.
This wandering amongst periodic (or, more generally, nonescaping)
orbits
results in the chaotic motion of passively advected particles.
Indeed, as long as the particles are in the mixing region,
their trajectories possess a positive average Lyapunov exponent $\lambda$.
Hence the union of all nonescaping orbits is called
the \emph{chaotic saddle}.
The flows relevant from our point of view
can be conisdered to be incompressible. This
results in a time-reversal invariant, area preserving
particle dynamics. Therefore, the negative average Lyapunov
exponent is exactly $-\lambda$, and it characterizes the
compression towards the chaotic saddle.

While many of the particles spend a long time
in the mixing region, the overwhelming majority of
them leaves this region
sooner or later.
The decay of their number in a fixed frame is
typically \emph{exponential}
with a positive exponent
$\kappa$ ($ < \lambda$), which is independent of the frame, i.e.,
$N(t) = N(0) \exp(-\kappa t)$.
This quantity $\kappa$
is the \emph{escape rate} from the saddle (or from the mixing region).
The reciprocal of the escape rate can 
be considered
as the average lifetime of chaos, and therefore the
chaotic advection of passive particles in open flows
is a kind of \emph{transient chaos} \citep{Tel90}.	

The chaotic saddle is the set of nonescaping orbits
which advected particles
may follow for an arbitrarily long time.
Each orbit of the set, and therefore the set as a whole,
has an  inflow  and an outflow curve, also called in the
mathematical jargon of chaos theory the stable and unstable
manifolds, respectively.
The \emph{inflow curve} is a set of points along which the
saddle can be reached after an infinitely long time.
The \emph{outflow curve} is the set along which
particles lying infinitesimally close to
the saddle will eventually leave it in the course of time.
By looking at different snapshots of these curves we can
observe that
they move periodically with the period $T$
of the flow. Their fractal dimension $D_0$ ($1< D_0 < 2$ in
two-dimensional flows) is, however, independent of the snapshot.
(The inflow and outflow curves
have identical fractal dimension due to the
advection dynamics' time reversal invariance.)

There is a unique relation between the fractal geometry and the
advection dynamics, expressed by the relation
\footnote{
By characterizing the dynamics by one single dimension $D_0$, we have assumed
that the advection process has a  monofractal geometry.
In reality, a set of dimensions $D_q$ is required for the
full description of the fractal aspects. It is for the $q=1$ dimension,
the so-called information dimension, $D_1$, for which
(\ref{KG}) is an exact equality.
In practice, however, the relative difference between $D_0$ and $D_1$ is
on the
order of a few  percents and therefore the use of a single dimension is
justified for practical purposes.}
\citep{Kan85,Hsu88,Tel90}:
\begin{equation}
D_0 = 2- \frac{\kappa}{\lambda}.
\label{KG}
\end{equation}
It says that the deviation of the dimension from that of the
plane is given by the ratio of two quantities characterizing the global
and the local instability of the dynamics. Relation (\ref{KG}) shows that
out of the three basic
characteristics ($\kappa$, $\lambda$ and $D_0$) only two are independent.
When speaking about population numbers in what follows, we shall use
the escape rate and the fractal dimension as independent parameters.
In the dynamics of the stripe widths (see next section), however, only the
average Lyapunov exponent appears.

The outflow curve plays a special role since it is the
only set which can be directly observed in an experiment.
Let us consider a droplet (ensemble)
of a large number of particles
which initially
overlaps with the inflow curve. As
the droplet is advected into the mixing region its
shape is strongly deformed, but the ensemble
comes closer and closer to the chaotic saddle as time goes on.
Since, however, only a small portion of particles can fall very
close to the inflow curve, the majority
does not reach the saddle and starts
flowing away from it along its outflow corve. Therefore,
in open flows  droplets of particles trace out
the outflow curve
of the chaotic saddle after a sufficiently long time of 
observation (in fact,
the populations in Fig.~\ref{fig:Compare} are distributed 
along the outflow curve
of the chaotic saddle present in the wake).

\section*{Dynamics of a single population}

In this section the mathematical derivation of
the dynamics of a single
population
living in an open  chaotic flow is briefly repeated 
\citep[][for more details see]{Tor98,Kar99,Tel00}.
Replication, competition for the limiting resources, and
spontaneous decay are taken into account in our population model,
while stage and age structure is neglected for simplicity.
We derive discrete and continuous-time models as well.

First we assume that the  intake of resource, multiplication and
decomposition are instantaneous and take place at integer
multiples of a time
lag $\tau$.
Here $\tau$ acts as an average time-scale on which the reproduction
takes place.

The basic observation is that after a sufficiently long time,
the filaments of the outflow curve are covered in narrow stripes
by individuals of species 
$B$  due to their replication \citep{Tor98,Kar99}. Individuals are thus
distributed on a fattened-up fractal
set. On linear scales larger than an average width $\varepsilon^*$ the
distribution of $B$
is a fractal of the same dimension $D_0$ as the outflow curve of
the chaotic saddle in the flow with passive advection
without biological activity. Let $\varepsilon^{(n)}$ 
denote the \emph{average} width
of  these stripes right before replication and decomposition takes place. It
is worth measuring this width in the unit of a characteristic  length scale
of the flow (e.g. in the cylinder radius in the example of 
Fig.~\ref{fig:Compare}). Thus,
$\varepsilon^{(n)}$ is a dimensionless variable.
Since material $A$ is
available outside of these stripes,
 replication increases the width with some
constant distance $\gamma$,
 the replication range,
 while spontaneous decay
due to
 death of individuals decreases it with a distance $\mu$. The  net
effect of
 the replication and spontaneous decay is then a broadening of the
width by an
 amount proportional to the  difference $\sigma=\gamma-\mu$, the
effective
 replication range. Thus,
$\varepsilon^{(n)} \rightarrow \varepsilon^{(n)} + c \sigma$. Here $c$ is a
dimensionless number expressing geometrical effects. If the fattened-up
filaments do not overlap, then replication does occur on the both side of
stripes and  $c=2$. If there is  overlapping among some of the fattened-up
filaments, such as in the case of a fractal,   then $c \neq 2$. This
\emph{geometrical factor} turns out to be slightly time dependent
due to the pulsation of the flow, but for
simplicity it can be considered to be constant from the point of view of the
qualitative behaviour of the population \citep{Tor01}.

In the next period of length $\tau$ there is no replication and decomposition,
just contraction
towards the outflow curve.
The average contraction factor is $\exp{(-\lambda \tau)}$, where $(-\lambda$)
is the negative average Lyapunov exponent of the advection dynamics. Therefore,
the
width $\varepsilon^{(n+1)}$ right before the next replication can be given as
\begin{equation} 
\varepsilon^{(n+1)}=(\varepsilon^{(n)} + c \sigma) e^{-\lambda
\tau}. 
\label{rece} 
\end{equation}  
This is a recursive map for the actual
width of the $B$-stripes on snapshots taken with
multiples of the time lag $\tau$. The solution of (\ref{rece}) converges
for $n \rightarrow \infty$ to the fixed point
\begin{equation}
\varepsilon^*= \frac{c \sigma}{ e^{\lambda\tau}-1}. \label{eps*}
\end{equation}

In the time-continuous
limit
$\tau \rightarrow 0$,
$\sigma \rightarrow 0$, but keeping $\sigma/\tau \equiv v_r$ constant,
one obtains
the differential equation:
\begin{equation}
\frac{d \varepsilon}{d t}=c v_r -\lambda \varepsilon,
\label{epsdiff}
\end{equation}
which has a steady-state solution given by:
\begin{equation}
\varepsilon^* = \frac{c v_r}{\lambda}\;. \label{ec*}
\end{equation}
Here $v_r$ can be interpreted as the  net speed of replication.

Knowing the $\varepsilon$-dynamics and
that the individuals accumulate on a fractal
set in the mixing
region, the time evolution of the number $N$ of $B$ individuals
in that region   can be
calculated. First, note that the area ${\cal A}$ occupied by species
$B$  scales as ${\cal A} \approx \epsilon^{2-D_0}$
with $D_0$ as the fractal dimension of the outflow curve for any box size
$\epsilon$ not smaller than the width $\varepsilon$ of the $B$-stripes. We can
thus choose\footnote{In general, (\ref{scale}) also contains a proportionality
constant, called the Hausdorff volume. Since this only rescales  the constant
$q$ in equation (\ref{discr}), for clarity we took the Hausdorff valume
to be unity.}
\begin {equation}
\epsilon=\varepsilon \approx {{\cal A}}^{1/(2-D_0)}. \label {scale}
\end {equation}
 If the linear size of the area occupied by a single individual is
$\epsilon_0$, we have $N = \epsilon_0^{-2} {\cal A}$, and therefore we can
rewrite (\ref {rece}) or (\ref{epsdiff}) so that it represents an equation for
the  individuals in discrete and continuous cases,
respectively:

\begin {equation}
N^{(n+1)}=e^{-\kappa \tau}
\{[N^{(n)}]^{1/(2-D_0)}+q \sigma\}^{(2-D_0)}, \label{discr}
\end {equation}
and
\begin{equation}
\frac{d N}{dt}=-\kappa
N+q (2-D_0) {v_r} N^{-\beta}, \label{diff}
\end{equation}
with
\begin{equation}
q = c \;\epsilon_0^{-2/(2-D_0)}. \label{eq:cn}
\end{equation}
 Here (\ref{KG}) has been used, and
\begin{equation}
\beta \equiv \frac{D_0-1}{2-D_0}
\label{beta}
\end{equation}
appears as a nontrivial exponent.%
\footnote{For multifractal flows
one can show
\citep{Tel00} that exponent $\beta$ is that given by
(\ref{beta}) with $D_0$ replaced by the information dimension
 $D_1$.}
Since the fractal dimension of the outflow curve lies
between $1$ and $2$, exponent $\beta$ is positive. For $D_0=1$ the differential
equation (\ref{diff}) describes a classical surface reaction along a line with
front velocity $v_r$ in the presence of escape. For $1<D_0<2$ it represents a
novel form of dynamical equations containing also an enhancing biological
activity term
with a negative power of the
area occupied by $B$ due to the fractality of the outflow curve.
The less $B$ individuals are present, the more
effective the reproduction is, because the resolved perimeter is
larger. Consequently, in a competitive situation the subordinate species has
an advantage if it becomes rare compared to the dominant species. This
balancing mechanism can make  coexistence possible, as shown in the
the next sections.

As one can see from Eqs.~(\ref{discr},\ref{diff}), in both the discrete
and continuum pictures a steady state is reached after a sufficiently
long time if the geometrical factor $c$ (and therefore also $q$)
is constant \citep{Tor01}. In this case, the steady-state number of
individuals in the mixing region is
$N^* =\epsilon_0^{-2}(\varepsilon^*)^{2-D_0}$ where
$\varepsilon^*$ is given by
(\ref{eps*}) and (\ref{ec*}) for the discrete and continuum cases,
respectively.

\section*{A model of competition}

As in the single species case, we consider a simple
model of replication and competition with
passively advected point like individuals
of type $B_1$ and $B_2$, multiplying themselves instantaneously.
The resource material $A$ which the different species $B_1$
and $B_2$ compete for
is uniformly distributed on the
surface of the flow.
Similarly to the single species case, the parameters $\gamma_i$ and $\mu_i$
($i=1,2$) are defined as the increase and decrease of the $B_i$ stripe width
due to replication and decomposition, respectively.
Therefore two
autocatalytic processes $A+B_1 \longrightarrow 2B_1$,
$B_1 \longrightarrow  A$ and $A+B_2 \longrightarrow 2B_2$,
$B_2 \longrightarrow A$
represent the  replication and competition
process in our model in an imperfectly mixed environment.
Similarly to the single species case, the parameters $\gamma_i$ and $\mu_i$
($i=1,2$) are defined as the increase and decrease of the $B_i$ stripe width
due to replication and decomposition, respectively, so that
the effective replication distances are
$\sigma_i=\gamma_i-\mu_i$.

As before, an important feature
of the advection dynamics is its  deterministic nature. Concerning the
population dynamics, this implies that  we work in the limit of weak diffusion
and assume that the mutual diffusion coefficients  between any pair of the
constituents is small.

Prior to discussing the consequences of the imperfect mixing generated by the
chaotic flow to this dynamics, it is worth briefly giving the
traditional equations governing the above defined autocatalytic processes
in a well-mixed environment. In a fixed region of observation
they are:
\begin {equation}
\frac{dN_1}{dt}=\gamma_1AN_1-\mu_1N_1,
\label {trad2}
\end {equation}
\begin {equation} 
\frac {dN_2}{dt}=\gamma_2AN_2-\mu_2N_2, 
\label {trad3}
\end {equation}
where $N_i$ denotes the number of individuals of
species $B_i$, and $A$ is the instantaneous
amount of the
resource material in the same region. Note that the meaning of the
replication and death rates are slightly different here from those in the  
discrete model (thus e.g., $\mu_i$ in eqs. (\ref{trad2}), (\ref{trad3})
is of dimension frequency, while the same quantity in the
discrete version is a distance).
If the dynamics of resource is much
faster than the dynamics of competing species, then
the former can be considered to be in a quasistationary state: $dA/dt=0$.
The equation for resource $A$ is then
\begin {equation}
\frac {dA}{dt} = 0 =l-\gamma_1AN_1 -\gamma_2AN_2, 
\label {trad1}
\end {equation}
and $l$ is the constant inflow of
resource $A$ into the region of observation.
Equations (\ref{trad2}), (\ref{trad3}) correspond to the general scheme
(\ref{trad0}) and (\ref{trad}) 
given in the Introduction by identifying $\mu_i$ with
$\delta_i$ and $\gamma_i A$ (where $A$ is given by the right hand side of
(\ref{trad1})) with $\alpha_i$. 

After analyzing (\ref {trad3}) and (\ref {trad1}),
one can easily see that species with lower ratio
$\gamma_i/\mu_i$ of replication and death rates
would be outcompeted, and thus stable coexistence is impossible.

\section* {Coexistence of competing species}

In the following, using a gedankenexperiment, we find
conditions
 under which the two species may coexist in the imperfectly mixing
environment
 characterized by the existence of a chaotic saddle and its
(fractal) outflow
 curve.
 First, let us assume that initially there is only
one species, for e.g., $B_2$
 in the mixing region. Also for the simplicity of
the writing, we will refer
 to the continuum version of the single species
dynamics, namely
 Eq.~(\ref{diff}).
 After a time the number $N_2$ of $B_2$
will be close to a steady state value,
 $N_{2}^*$. We now let a small quantity
of species $B_1$ invade the
 mixing region, so small
 that it cannot change
the steady state of $B_2$.
 The question is: under what conditions this
invasion
 will lead to a self-sustained $B_1$ population, coexisting with
$B_2$?
 
When we are letting species
$B_1$ invade the mixing region such that $B_2$ is already present, we must
ensure for the coexistence that there is not only $B_1 - B_2$ interface
present but also $B_1 - A$ interface, at all times. To
show that this is indeed  possible, we recall the Wada property of mixing on
the chaotic saddle,  mentioned previously. As shown by 
\citep{Tor97},
if species $B_1$ is inserted such that the $B_1 - A$ interface
\emph{intersects} 
the inflow curve, then this boundary must
be present arbitrarily close to all points on the outflow curve. Therefore
this initial condition ensures the existence of the $B_1 - A$ interface
at all times on the saddle and its outflow curve.
Thus, when there are interactions among the species,
then, due to the relatively long time they spend on this
fractal, and due to the largely increased interfaces
a non-trivial and novel type of behavior can emerge.

Species $B_1$  becomes thus trapped by		
the chaotic saddle, and will be distributed in very narrow filaments
along the surface  of some of the $B_2$ stripes close to the
$D_0$ dimensional outflow curve.
Since by assumption, the population number $N_1$ of species
$B_1$ is small, ${N_{1}}/{N_{2}} \ll 1$, there is no feedback on $B_2$
and this species remains in dynamical equilibrium:		
${d N_{2}}/{d t}\approx 0$.
The dynamics of species $B_1$ can thus
be described by (\ref{diff}) written for $N_1$.
The
important nature of (\ref{diff}) is that the first term on the right side,
(which is responsible for the outflow from the fractal set)
tends to zero
for very small $N_{1}$. The second term, however,
which describes an autocatalytic process can be arbitrarily large if
$N_{1}$ is small,
due to the negative exponent $-\beta$.
Thus, if $N_{1}$ is sufficiently small then  ${d
N_{1}}/{d t} > 0$.
It means that the $B_1$ population always increases if it's
number is close to zero.
In other words, the fixed point $N_1^*=0$ is unstable.


Following the same argument,
if ${B_1}$ and ${B_2}$ are already in dynamical
equilibrium on the fractal set then a third species can invade this coalition
in the same manner.
This coexistence has indeed been demonstrated for
three species by numerical simulations \citep{Kar00}.


In this gedankenexperiment we assumed that the invading population
is so weak that it is distributed in narrow filaments
along the ``surface'' of the already existing population and does not
influence this population at all. After the number of the invading
population has started to grow, there is an increasing interaction
between the populations due to the competition for the same
resource. As a consequence, the originally steady population
is not
in a stationary state and the dynamics
should be treated in a selfconsistent manner. This is
shown in the next section.

\section*{A mathematical model for the competition dynamics}

After sufficiently long time,
both species $B_1$ and $B_2$  will be distributed in
narrow stripes along the chaotic saddle's outflow curve as follows from the
passive advection dynamics. Due to the replication and decomposition,
however, the stripes have finite widths (cf.\ Fig.~\ref{fig:Compare}) 
which might depend on time.
Let $\varepsilon^{n}$ denote the dimensionless average width of the stripes
right before an instantaneous replication
takes place. These stripes are
defined by the fact that outside of them there is only background material
$A$ available. Inside the stripe of width
$\varepsilon^{(n)}$
there might be several narrow $B_1$ or $B_2$
filaments.
The background material $A$
is eaten up sooner or later in the inside of any stripe, therefore,
for the sake of an easier presentation,  we assume that this is the case and
only material $B_1$ and $B_2$ are present. Let us
denote the total widths of all the filaments of a given material  within an
$\varepsilon^{(n)}$ stripe  by $\varepsilon_i^{(n)}$
with $i=1,2$ corresponding to $B_1$ and $B_2$, respectively.
 The sum of
these partial widths is of course the total one
$\varepsilon_1^{(n)} + \varepsilon_2^{(n)}=
\varepsilon^{(n)}$.
Our aim is to build up the dynamics of the partial widths
based on plausible assumptions, from which
the dynamics of the different populations follows.

We assume,
that the boundaries are occupied  by
species $B_1$ or $B_2$
with \emph{probabilities} $p_1$ and $p_2$, respectively.
In other words, a stripe-boundary picked at random from
the many filaments of the outflow
curve will have a probability  $p_i$ to be of type $B_i$,
$i=1,2$.
If mixing of the two species were
perfect along the fractal set, these probabilities would be equal to their
relative densities.
This is not the case, however.
The relative position of the species in the initial
distribution  to the inflow
curve determines which individual or patch of individuals will be
trapped by which orbit of the chaotic saddle.
The rest, i.e., the untrapped individuals will drift out of the
mixing region.
The trapped individuals, however, will stay there forever, and follow
their specific trapped orbit.
In the course of time, individuals give birth to others of the
same species, and patches of individuals are stretched along the
outflow curve specific to the trapping orbit of the chaotic saddle.
In either cases, we end up with long stripes of the two species
lined up along each other in an alternating manner, tracing out the
outflow curve.
Then the probability of one species to be on the edge of these lines,
and thus to be capable of reproduction, depends
on which trapping orbit will produce the filament of
outflow curve being on the edge of the stripe,
on the order in which the species are lined up across one stripe,
and on the actual width of the coverage of the filaments.
In other words, it is the complex chaotic dynamics which makes
the introduction of probabilistic concepts---on a somewhat
phenomenological level---unavoidable.

The probabilities $p_i$
depend on what the distribution of the species
inside the stripes is. 
Thus, the simplest possible assumption is that the probabilities
depend on the partial widths $\varepsilon_i^{(n)}$.
Their actual functional
form might also contain parameters of the flow and of the
biological activity.

Naturally the probabilities fulfill $0 \le p_1^{(n)} \le 1$ and
$p_2^{(n)}=1-p_1^{(n)}$.
They might have a general dependence
on the partial widths $\varepsilon_i^{(n)}$,
$i=1,2$: $p_i^{(n)}=p_i^{(n)}\left(\varepsilon_1^{(n)},\varepsilon_2^{(n)}
\right)$.
Due to dimensional reasons they can only depend on the
ratio $z^{(n)}\equiv \varepsilon_1^{(n)}/\varepsilon_2^{(n)}$. Thus
we write
\begin{equation}
p_1^{(n)} = g\left(z^{(n)}, \omega \right), \;\;\;\;\;
p_2^{(n)}=1-g\left(z^{(n)}, \omega \right). \label{pe}
\end{equation}
Here $\omega$ is a parameter of the distributions, and incorporates
the dependence on the replication rates.
We also made the plausible assumption, that $g$ has no explicit $n$ (or
time) dependence.
A general property of $g$ is that it vanishes in the origin
$g(0) = 0$ since this expresses the obvious fact that
if species $B_1$ is missing, then the probability to find it
in the filaments is zero. Similarly for infinitely large values of
$z$ it must be unity: $g(\infty)=1$ which corresponds to the absence
of $B_2$.
Also, due to the fact that $g(z)$ is a
probability distribution, we must have $0 \leq g(z) \leq 1$
for all $z \geq 0$. Furthermore, the functional form must be
symmetric by interchanging the role of the species. This implies
that one must have $g'(0) \geq 0$, where the prime denotes
derivation with respect to the argument.
This implies
\begin{equation}
p_2^{(n)}=g\left(1/z^{(n)},1/\omega \right),
\end{equation}
where the appearance of $1/\omega$ means that an interchange of the
species index brings the parameter in its reciprocal value,
as e.g.\ in the case when $\omega=\sigma_1/\sigma_2$ (the dependence 
on the ratio of the replication distances follows from dimensional
reasons).
The normalization of the probability implies
\begin{equation}
g\left(z,\omega \right)+
g\left(1/z,1/\omega \right)=1 \label{feltetel}
\end{equation}
This is a functional equation for $g$. With the above properties and
boundary conditions we find that a family of solutions is given by
the form:
\begin{equation}
g(z) = \frac{z^{\alpha}}{z^{\alpha} +\omega}, \label{pe2}
\end{equation}
with $\alpha$ and $\omega$ as two positive parameters. In the range
of $0<\alpha<1$ the smaller population is less probable on the
boundary but yet with a weight which is weaker than linear in the widths.
For $\alpha=0$ there is no width-dependence at all, the probabilities
$p_i$ are constant.
The case $\alpha=1$ and $\omega=1$ corresponds to a homogeneous mixing within
the stripe of width $\varepsilon$. For $\alpha>1$ a superdominance is
described.
In the next section we show that  the form
(\ref{pe2}) of $g(z)$ is indeed in good agreement with
numerical simulations, and determine values for parameters
$\alpha$ and $\omega$.

The broadening of the average widths is then $c \sigma_1 p_1^{(n)}$ and
$c \sigma_2 p_2^{(n)}$  due to species $B_1$ and $B_2$, respectively. Here
the geometrical factor $c$
and parameter $\sigma_i=\gamma_i-\mu_i$ have the same meanings  as
in the single species problem defined previously.

Thus, similar to (\ref{rece}) the partial width of
$B_i$
after the $(n+1)$st
step
is
\begin{equation}
{\varepsilon_i^{(n+1)}}=\left[
{\varepsilon_i^{(n)}}+
 c \sigma_i p_i^{(n)} \right] e^{-\lambda \tau}
\label{eq:epsi}
\end{equation}
for $i=1,2$.
Note that in our theory, $c p_1$ and $c p_2$
can also be interpreted as renormalized
geometric factors for each species, due to the screening effects at the
boundaries of the stripes.
As a consequence, the total width of the stripes changes at a replication
as
\begin{equation}
{\varepsilon^{(n+1)}}=\left[
{\varepsilon^{(n)}}+ c
( \sigma_1 p_1^{(n)}+ \sigma_2 p_2^{(n)})
\right]e^{-\lambda \tau}.
\label{eq:eps}
\end{equation}
For simplicity, the explicit width-dependence (\ref{pe}) 
of the probabilities
has not been written out.
For 
$\sigma_1=\sigma_2$ we recover the width dynamics
of the single species problem, see (\ref{rece}).

Next we turn to the dynamics of the number of individuals.
On scales larger than or equal to $\varepsilon^{(n)}$,
the total  number of individuals $N =N_1 + N_2$
occupied by stripes appears to be a fractal of the same
dimension $D_0$ as the outflow curve. For simplicity of writing we assume
that both species have the same size $\varepsilon_0$ (an extension for
different sizes is straightforward).

Since the relation between the $\varepsilon^{(n)}$ and the number of
individuals $N^{(n)}$ is the same as in the single species
model, we can use (\ref {scale}).
Thus (\ref{eq:eps})
implies a recursion  for the  area
right before  replication as
\begin{equation}
N^{(n+1)}\!\!= \!e^{-\kappa \tau}\!\!
\left\{ \!\left[ N^{(n)}\right]^{1/(2-D_0)} \! +\!
q\left[\sigma_1 p_1^{(n)} + \sigma_2 p_2^{(n)} \right]
\right\}^{2-D_0}\label{eq:mapauto}
\end{equation}
with $q$ given by (\ref{eq:cn}).
Next, we derive the dynamics of the  number of
individuals $N_i^{(n)}$ for 
species $i$ contained in the stripes.
The number of individuals of species $i$ is the 
portion of the total
number $N^{(n)}$ proportional to the partial widths:
\begin{equation}
N_i^{(n)}=N^{(n)}
\frac{\varepsilon_i^{(n)}}{\varepsilon^{(n)}}. \label{eq:Ai}
\end{equation}
This is due to the fact that there is no
fractal scaling below $\varepsilon^{(n)}$. Since
$\varepsilon^{(n)}=[\epsilon_0^2 N^{(n)}]^{1/(2-D_0)}$,
Eq.~(\ref{eq:Ai}) leads to
\begin{equation}
\varepsilon_i^{(n)} = N_i^{(n)} \epsilon_0^{2/(2-D_0)}
\left[ N^{(n)} \right]^{\beta}. \label{aiei}
\end{equation}
As another consequence of (\ref{eq:Ai}), the ratio of the partial widths
is the ratio of the population numbers:
\begin{equation}
z^{(n)} \equiv \frac{\varepsilon_1^{(n)}}{\varepsilon_2^{(n)}} =
\frac{N_1^{(n)}}{N_2^{(n)}}.
\label{pA}
\end{equation}
{From} Eqs.~(\ref{eq:epsi}) and (\ref{aiei})
we therefore  obtain the dynamics of the population
numbers as
\begin{eqnarray}
N_i^{(n+1)}[N^{(n+1)}]^\beta = e^{-\lambda\tau}
\left\{ N_i^{(n)}\left[N^{(n)}\right]^\beta 
\right. && \nonumber \\
\left. +  q \sigma_i p_i^{(n)}
\left( N_1^{(n)}/
N_2^{(n)}\right) \right\} && \label{eq:mapautoA}
\end{eqnarray}
for $i=1,2$. Here exponent $\beta$ is the same expression
(\ref{beta}) as in the case of the single species problem,
and $q$ is given by (\ref{eq:cn}).

We observe that
by dividing the rearranged
(\ref{eq:mapautoA}) for $i=1$ by that with $i=2$, one obtains
\begin{equation}
\frac{\sigma_1 p^{(n)}_1}{\sigma_2 p^{(n)}_2}= M^{(n)},
\end{equation}
where
\begin{equation}
M^{(n)}=\frac{e^{\lambda}N_1^{(n+1)}[N^{(n+1)}]^{\beta}-
N_1^{(n)}[N^{(n)}]^{\beta}}{
e^{\lambda}N_2^{(n+1)}[N^{(n+1)}]^{\beta}-
N_2^{(n)}[N^{(n)}]^{\beta}}.
\end{equation}
{From} this, $p_1=1-p_2$ is easily found as
\begin{equation}
p^{(n)}_1=\frac{M^{(n)}}{M^{(n)}+
\frac{\sigma_1}{\sigma_2}},\label{eqfit}
\end{equation}
This relation provides us with a method for measuring
how the probability $p_1(z) \equiv g(z)$ depends on the
ratio $z\equiv N_1/N_2$ at any instant of time. We shall use
this observation to extract the form of the $g$ function from
numerical results. The ratio of fixed points,
$z^* \equiv \frac{N_1^*}{N_2^*}$ can be calculated from (\ref{eqfit})
assuming that $N_2^* \neq 0$ as
\begin {equation}
g(z^*)=\frac{M^{*}}{M^{*}+\frac{\sigma_1}{\sigma_2}},\label{fixdisc}
\end{equation}

The time continuous limit is obtained
by letting  both the time lag and the effective replication
ranges go to zero so that their
ratios remain finite. Thus we define replication velocities
\begin{equation}
v_i = \lim_{\tau \rightarrow 0} \frac{\sigma_i}{\tau},
\end{equation}
with $i=1,2$ for species $B_1$, $B_2$, respectively.
In the continuous time limit,
the differential equations
obtained for the partial widths from (\ref{eq:epsi}) read as
\begin{equation}
\frac{d \varepsilon_i}{dt}=-\lambda {{\varepsilon}_i}
+ c v_i p_i(\varepsilon_1 / \varepsilon_2)
\label{epsi}
\end{equation}
where $p_1=g$, $p_2=1-g$.

The differential equation for the number of all individuals follows from
(\ref{eq:mapauto}) as
\begin{equation}
\frac{d N}{dt} = -\kappa N+ q (2-D_0) v N^{-\beta}.
\label{diffeq}
\end{equation}
Here
\begin{equation}
v \equiv p_1 v_1 + p_2 v_2
\label{avv}
\end{equation}
is an average velocity, but note that it is not a constant since
the $p_i$ depend on the population numbers, and $q$ is given by (\ref{eq:cn}).

The differential equation for the number 
$N_i$ of individuals of
the two species can be derived from (\ref{epsi}) and the continuum
version of (\ref{aiei}), i.e., 
$\varepsilon_i = \epsilon_0^{2/(2-D_0)}N_i
N^{\beta}$. The result is
\begin{eqnarray}
\frac{d N_i}{dt} = -\kappa N_i-q(D_0-1) v
N^{-\beta-1} N_i+ &&\nonumber \\
q v_i p_i \left( N_1 / N_2\right) N^{-\beta},&& \label{diffeqi}
\end{eqnarray}
with $N = N_1 + N_2$.
Here (\ref{KG}) and (\ref{beta}) have been used.
By summing over $i$ in (\ref{diffeqi})
one recovers Eq.~(\ref{diffeq}).

An equivalent form is obtained after rearranging terms and 
taking into account the
definition of the average replication velocity (\ref{avv}). It reads
\begin{eqnarray}
\frac{d N_1}{dt} = -\kappa N_1 +
qN^{-\beta-1}\left[ (2-D_0)v N_1+ \right. &&\nonumber \\
\left.(v_1 p_1 N_2 -v_2 p_2 N_1) \right] &&,\label{diffeq1}
\end{eqnarray}
and an analogous expression for the second species obtained 
from (\ref{diffeq1}) by interchanging the indices $1$ and $2$.
It can be clearly seen that the first term of the
bracket corresponds to the growth 
of the total population, while the second describes the effect due to a
weighted difference in the population numbers.
Expression (\ref{diffeqi}) or (\ref{diffeq1})
represents a strongly coupled set of nonlinear equations
with  a novel type of power-law behavior (with negative exponent
$-\beta$). This set of equations is the central result of our paper since it can 
be
considered as a population dynamics
describing the coupling of two  populations {mixing} on a fractal,
and as we show below,
opens up the possibility to have a nontrivial
coexistence.

If one of the species, say $B_2$, is not present, then $p_1=1$, 
$N_2=p_2=v_2=0$ and
hence $v=v_1=v_r$, $N=N_1$ and Eq.~(\ref{diffeq}) becomes equivalent to
(\ref{diff}). The same happens if both species are equivalent, i.e., for
$v_1=v_2$ when $v=v_r$. 

A simple further equivalent  form can be derived
by using relative densities $c_i \equiv N_i/N$. 
The equations describing the populations then become
(by using (\ref{diffeq1}) and (\ref{diffeq}):
\begin{equation}
\frac{d c_1}{dt}  = q N^{-\beta - 1} (v_1 p_1 c_2- v_2 p_2 c_1) 
\end{equation}
with $c_1+c_2=1$. The temporal change of the densities is determined
by the weighted relative difference in the densities.
 Note that they  are {\em multiplicatively}
coupled to $N^{-\beta -1}$ which is proportional to the average
width of the filament covering. For $D_0 = 1$ this is just $1/N$ and
it is the { spatial concentration} or the density of the total population.
For $2 > D_0 > 1$ (fractals), this factor is the \emph{fractal} 
spatial density of the population as a whole.

\section*{Coexistence analysis: fixed points and their stabilities}

It is simple and instructive to study the time-continuous
dynamics of the widths
$\varepsilon_i$, $i=1,2$ in and around steady states.
Assuming, as before,  $c=\mbox{const.}$, we find from (\ref{epsi})
\begin{equation}
\lambda {{\varepsilon}_i^*}=
c v_i p_i \left( {\varepsilon_1^*} 
/ {\varepsilon_2^*} \right).\label{fpi} 
\end{equation}
The weighted sum of the fixed points widths (\ref{fpi}) gives
\begin{equation}
\varepsilon_1^* v_2+
\varepsilon_2^* v_1=  \frac{c v_1 v_2}{ \lambda}. \label{fp1}
\end{equation}
{From} here on, for the stability analysis, we shall use the
explicit form (\ref{pe2}) for the function $g$.
Thus (\ref{fpi}) translates into:
\begin{eqnarray}
&&\lambda {{\varepsilon}_1^*}=
c v_1 \frac{ {\varepsilon_1^*}^{\alpha}}{
{\varepsilon_1^*}^{\alpha}+\omega
{\varepsilon_2^*}^{\alpha}} \label{fpio1} \\
&& \lambda {{\varepsilon}_2^*}=
c v_2  \frac{ {\omega\varepsilon_2^*}^{\alpha}}{
{\varepsilon_1^*}^{\alpha}+\omega
{\varepsilon_2^*}^{\alpha}} \label{fpio2}
\end{eqnarray}

Formula (\ref{fp1}) shows that one species always survives. Without
loss of generality, we can choose this to be species $B_2$, and
everything remains valid with the indices 1 and 2 switched.
It is worth defining:
\begin{equation}
z^* \equiv \varepsilon_1^*/\varepsilon_2^*.  \label{zst}
\end{equation}
Since $\varepsilon_2^*$ is not zero, the ratio of the fixed point equations
(\ref{fpio1},\ref{fpio2}) yields:
\begin{equation}
{z^*}^{1-\alpha} = \frac{v_1}{v_2 \omega}\;\;\;\mbox{or}\;\;\;
{z^*} = 0\label{fp2}
\end{equation}
The first equality describes the $z^* \neq 0$ \emph{coexistence}
fixedpoint while the second describes the \emph{non-coexistence} fixedpoint.

The  equations (\ref{epsi}) of the continuous case  written out 
explicitely are as follows:
\begin{eqnarray}
&&\frac{d \varepsilon_1}{dt}=-\lambda {{\varepsilon}_1}
+ c v_1 \frac{\varepsilon_1^{\alpha}}{\varepsilon_1^{\alpha} + 
\omega \varepsilon_2^{\alpha}} \label{flow2} \\
&&\frac{d {\varepsilon}_2}{dt}=-\lambda {{\varepsilon}_2}
+ c v_2 \frac{\omega \varepsilon_2^{\alpha}}{\varepsilon_1^{\alpha} + 
\omega \varepsilon_2^{\alpha}} \label{flow1}
\end{eqnarray}
The linear stability of a fixed point  $(\varepsilon_1^{*}>0,\varepsilon_2^{*}>0)$
will be
given by the eigenvalues of the stability matrix ${\bf E}$, calculated from
(\ref{flow2}-\ref{flow1}) as follows (here we also used (\ref{fpio1}-\ref {fpio2}):
\begin{eqnarray}
{\bf E} \!= \!\lambda \left( \!\!
\begin{array}{ll}
\! -1 +\frac{\alpha \lambda}{c v_2}\varepsilon_2^* & \;\;
-\frac{\alpha \lambda}{c v_2}\varepsilon_1^* \\
& \\
\! -\frac {\alpha \lambda}{c v_1} \varepsilon_2^* & \!\!
-1 + \frac{\alpha \lambda}{c v_1}\varepsilon_1^*
\end{array}\!\!
\right) \nonumber
\end{eqnarray}
The eigenvalues
of ${\bf E}$ are easily calculated:

\begin{equation}
\Lambda_+=-\lambda (1-\alpha),\quad \Lambda_-=-\lambda . \label{coex}
\end{equation}

One eigenvalue of the width dynamics is always the negative of the chaotic
advection's positive Lyapunov exponent. As long as the parameter $\alpha$ is
less than unity, the other eigenvalue is also negative.

We find that for
$0< \alpha < 1$ \emph{coexistence is stable}, for $\alpha > 1$ it 
becomes unstable.  

The case $\alpha=1$ is special.
It follows from (\ref {fp2}) that
for $\alpha=1$, and $\omega \neq v_1/v_2$, the
non-coexsitence point is the only fixed point, and it is stable.
(If $\omega > v_1/ v_2$ then 
($\varepsilon_1^* > 0, \varepsilon_2^* = 0$) is stable, if $\omega < v_1/ v_2$ then 
($\varepsilon_1^* = 0, \varepsilon_2^* > 0$) is stable fixed point.)
   Having
$\alpha=1$  with $\omega= v_1/v_2$ implies that all 
points fulfilling (\ref{fp1})
are fixed points of marginal stability.
Thus, stable coexistence
is found in the 
\begin{equation}
0 < \alpha < 1
\end{equation}
regime.

Interestingly, the stability conditions are the same for
discrete-time dynamics, as well.

In the next section we analyze coexistence in a simple
chaotic dynamical system, the Baker map,
where we show that  $g(z)$ is given by Eq.~(\ref{pe2})
in this process,
and determine the parameters $\alpha$ and $\omega$ from
numerical experiments.

\section*{Numerical results}

In this section we present numerical verification of the
new type of population dynamical equation we introduced before.
We have already shown that coexistence in open
flows is possible \citep{Sch00b,Kar00},
so we deal only  with the  \emph{quantitative} verification
of the theoretical results.

\begin{figure}[htbp]
\vspace*{-0.5cm}
\hspace*{-0.5cm}
\epsfxsize=3.4 in \epsfbox{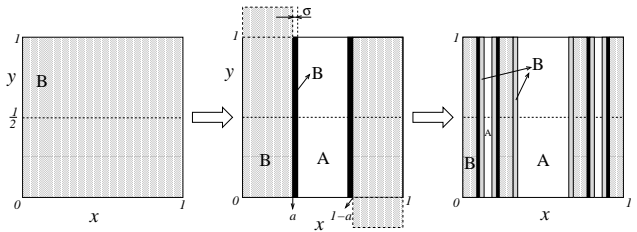}
\vspace*{0.2cm}
\caption{Two consecutive steps of the baker map and two replications
($\tau=1$) for the single species model.
The bands of width $\sigma$ become occupied by $B$ in each replication.
The material hanging over the unit square is discarded.
}\label{figbaker}
\end{figure}

For computational simplicity, we use the so-called
\emph{baker map} to model the flow \citep{Tor01}.
This can be considered as a simplified model of
stretching and folding in a chaotic flow observed
periodically after specified time-intervals.
Thus, in this case $\tau$, the time lag between instantaneos
multiplications of the species, is an integer number denoting
the number of snapshots taken of the flow between two
consequtive multiplications.
The baker map,
acting on the unit square,
gives the new location $(x',y')$ of an individual starting at
point $(x,y)$:
\begin{eqnarray}
\begin{array}{ll}
x'=ax+(1-a)\theta(y-1/2), & x\in[0,1],\\
y'=\frac{1}{a} y-(\frac{1}{a}-1)\theta(y-1/2), & y\in[0,1],
\end{array}
\end{eqnarray}
where $a<1/2$ is the parameter of the baker map,
and $\theta(x)$ is the Heaviside step function.
The action of the baker map is shown schematically in
Fig.~\ref{figbaker}.
The area preserving property of this baker map models
the incompressibility of realistic hydrodynamical flows,
while outflow is modelled by neglecting the area hanging over
the edge of the unit square.

Starting with any initial conditions, after a few steps of
iterations both species will be distributed along
narrow filaments parallel  to the $y$ axis.

\begin{figure}[htbp]
\vspace*{-0.5cm}
\protect\hspace*{-0.5cm}
\epsfxsize=3.2 in
\epsfbox{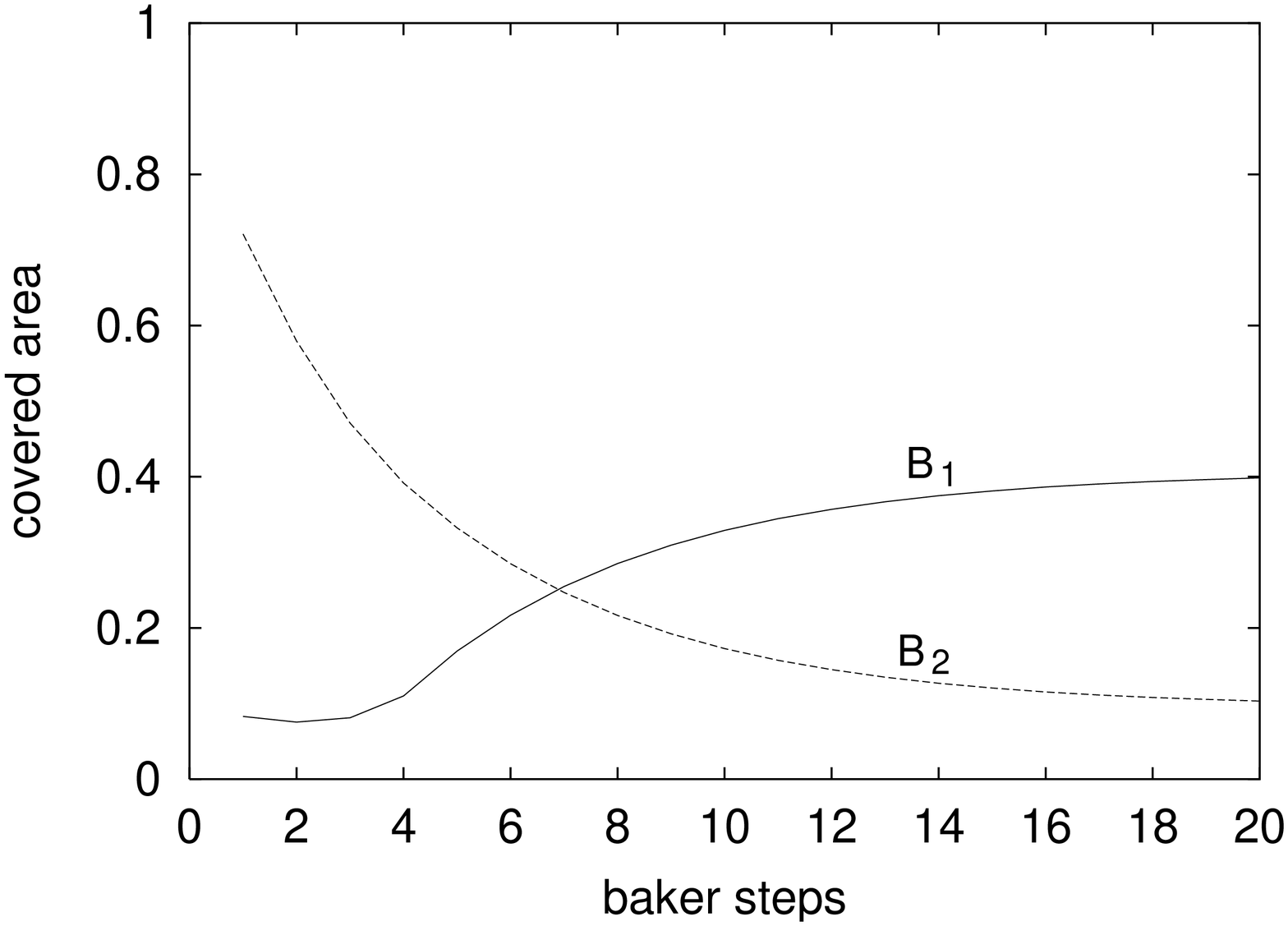}
\protect\vspace*{0.2cm}
\caption{Reaching the equilibrium states with the coexistence of
two species is shown. Initially, two patches of species were
placed, one patch of $B_1$ in $x\in [0; 0.1]$, $y\in [0;1]$,
and another
patch of
$B_2$ at $x\in [0.1; 1]$, $y\in [0;1]$.
The parameter values are $a=0.4$ 
for the baker map, and $\sigma_1=0.003$
$\sigma_2=0.001$ for the competing species.
The areas covered by the species is shown right after
the multiplications
taking place.
After an initial transient (time-steps 1--4), we have a rapid
convergence to the fixed point (time-steps 5--18), after that
we have
an equilibrium setting in (time-steps 18--20).
}
\label{fig:timeevol}
\end{figure}

\begin{figure}
\protect\hspace*{-0.5cm}
\epsfxsize=3.2 in
\epsfbox{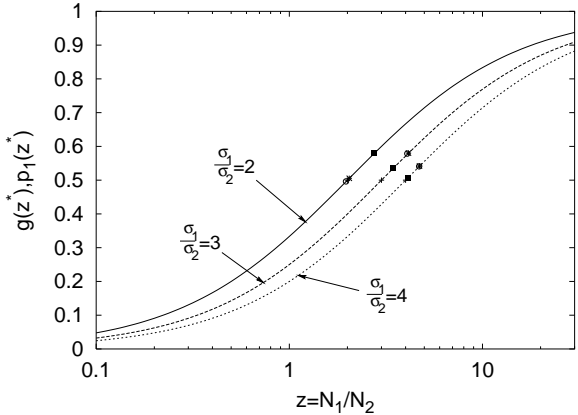}
\protect\vspace*{0.2cm}
\caption{The dependence of the probabilities $p_1$
on $N_1/N_2$ in the nontrivial fixed points.
The initial  positions do not influence the fixed point reached.
The curve
$g(z)=z/(z+\sigma_1/\sigma_2)$ is
shown with
solid line for $\sigma_1=0.002$, $\sigma_2=0.001$,
with dashed line for $\sigma_1=0.003$, $\sigma_2=0.001$,
and with dotted line for $\sigma_1=0.004$, $\sigma_2=0.001$.
All the measured fixed point values  fulfill
$g(z^*)=z^*/(z^*+\sigma_1/\sigma_2)$.
The fixed points are marked by crosses ($a=0.25$), black
squares ($a=0.3$),
starts ($a=0.35$), and circles ($a=0.4$ as the baker parameter).
}
\label{fig:fixedpoint}
\end{figure}

After $\tau$ baker steps, individuals of species $B_i$
multiply and give birth into a vertical stripe of width
$\sigma_i$ covered by resource $A$,
lying along the borderline of the previously
occupied region of species $B_i$ parallel to the $y$ axis.
In the numerical experiments, we used $\tau=1$, that is,
the species multiplied after each baker-step.
Regions which are invaded by both species after
instantaneous multiplication are divided between them in a ratio
of $\sigma_1/\sigma_2$.
It is expected that (\ref{eq:mapautoA})
describes the time-evolution
of the species, reaching the fixed-point (\ref{fixdisc}) eventually.
Figure~\ref{fig:timeevol} shows in a typical case how the
equilibrium state with coexistence is reached after about
18 baker steps.
Similar results were obtained with various other parameter
settings, in accordance with the theoretical results.
%
We checked the validity of the form
(\ref{fixdisc}) against the
numerical results in steady states.
$N_i^*$ is the fixed-point number of individuals of species
$B_i$. The fixed point values are found to fulfill
(\ref{fixdisc}), see
Fig.~\ref{fig:fixedpoint}.

\begin{figure}
\protect\hspace*{-0.5cm}
\epsfxsize=3.2 in
\epsfbox{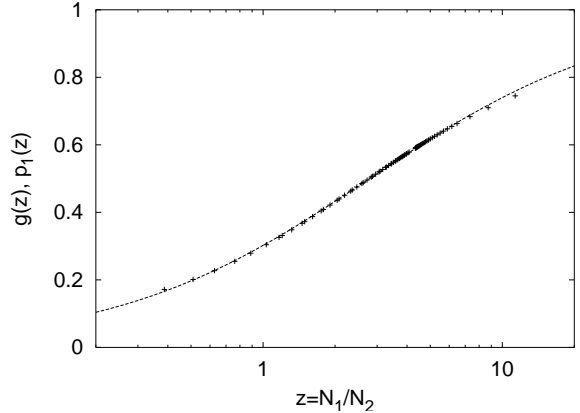}
\protect\vspace*{0.2cm}
\caption{The dependence of the probabilities $p_1$
on $N_1/N_2$
with $\sigma_1=0.003$, $\sigma_2=0.001$ values, and with five
different initial conditions.
The parameter of the baker map was $a=0.4$.
The initial positions do not influence the fixed point reached.
Solid line shows the function $g(z)=z^{\alpha}/(z^{\alpha}+\omega)$
with $\alpha=0.818\pm 0.002$ and $\omega=2.312\pm 0.006$.
The parameter of the baker map was $a=0.4$.
}
\label{fig:figfit}
\end{figure}

Next we check
the validity of (\ref{eq:mapautoA}) for the time-evolution
before reaching the convergence. We measure the population numbers
in discrete time $n$ and
use relation (\ref{eqfit})   to extract the form of the
probability distribution $g$.
Figure~\ref{fig:figfit} shows $p_1$ as a function of $N_1/N_2$
for fixed parameter values,
but for various initial conditions.
There is a single function covering the measured points
which can well be fitted by the form
$g_1(z)=z^{\alpha}/(z^{\alpha}+\omega$.
In all cases $\alpha<1$ was measured indicating that the coexistence
fixed point is stable.
Also note that $\omega\approx (\sigma_1/\sigma_2)^{\alpha}$
was found in all experiments, which implies that (\ref{feltetel}) holds.

\begin{figure}
\protect\hspace*{-0.5cm}
\epsfxsize=3.2 in
\epsfbox{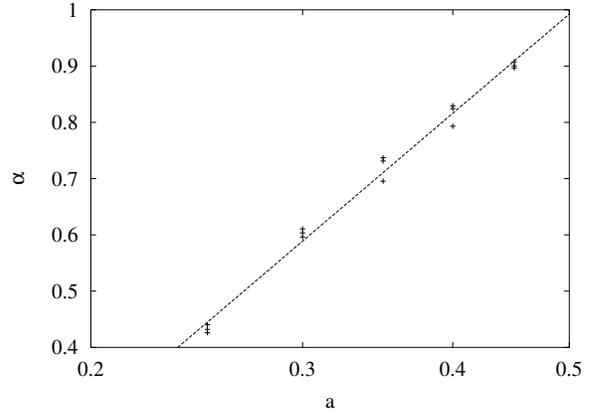}
\protect\vspace*{0.2cm}
\caption{
Dependence of $\alpha$ on the parameter $a$ of the baker map.
Various points are measured values for multiple $\sigma_1/\sigma_2$
ratios. The dashed line is the function $\alpha(a)=0.79\ln a+1.54$.}
\label{fig:fit}
\end{figure}

We also measured how $\alpha$ depends on the
parameter of the baker map,
or, on the fractal dimension $D_0$ of the
outflow curve of the chaotic
saddle.
We found that $\alpha=0.79\ln a+1.54$, see Fig.~\ref{fig:fit}.
Using the fact that $D_0=\ln 2/ln(1/a)$, we obtain
$\alpha=1.54-0.55/D_0$.

\section* {Discussion}

We derived a novel type of coupled population dynamic equations for
two populations competing on a fractal set provided by open chaotic flows.
The equation for the number of species in a given fixed range
of the flow can be written in the general scheme (cf.\ \ref{diffeq1})
\begin{equation}
\frac{dN_1}{dt} = \alpha_1\left(\frac{N_1}{N}\right) N_1^{-\beta} - 
\kappa N_1,
\end{equation}
\begin{equation}
\frac{dN_2}{dt} = \alpha_2\left(\frac{N_2}{N}\right) N_2^{-\beta} - 
\kappa N_2.
\end{equation}
The coefficients $\alpha_i$ of the replication terms depend on the relative
densitites (denoted by $N_i/N \equiv c_i$) only.
Their explicit form follows from  (\ref{diffeq1}). For example,
\begin{equation}
 \alpha_1\left(\frac{N_1}{N}\right) =
 q \left( \frac{N_1}{N} \right)^{\beta} \left[(1-D_0) v \frac{N_1}{N}- v_1 p_1
\right].
\end{equation} 
The structure of these equations is similar to that of (\ref{trad0}),
(\ref {trad}) or
(\ref{trad2}, \ref {trad3}). The time derivative is the sum of a gain term
 and a
loss term, but now the gain term contains a nontrivial
 negative power of the
population number and is coupled
 to the other population in a nonlinear way.
These equations
 describe  the population dynamics in an imperfectly mixed
enviroment of dimension $1<D_0<2$. The fractality $D_0$
of the mixing region (in our case of the outflow curve)
appears in the power
$\beta=(2-D_0)/(D_0-1)$.
In this set of equations a phenomenological
function ($p_1$) is also present characterizing the probability that a given
population is on the surface of the fractal support
with free access to the single
available resource. Based on general arguments and a simple model,
this function turned out to be a normalized power law distribution
of the type of (\ref{pe2}).
This form
expresses a kind of ``advantage of rarity'' principle:
for exponent $0<\alpha < 1$ the derivative is infinite in the origin,
a very small increase in the size of the weaker population
leads to a drastical increase of the probability for being on
the free surface and hence to grow.
On the contrary, for $\alpha>1$,
only a relatively large
population size has considerable growing probability, in this case
the weaker population dies out. It is worth mentioning that in
the gedankenexperiment of Section ``Coexistence of competing species'', 
we did not see this effect since
we had an initial stage at which the weaker population
does not yet influence the stronger one. The probability of being
on the surface was assumed to be constant. It is the interaction
between the two populations which leads to the power law distribution.
Its exponent is determined by the flow and the biological
process.
With exponents larger than unity this form does not allow for
coexistence.
The presented mathematical forms and the conditions
for coexistence remain valid if $m >2$ species live in
open chaotic flow,
a numerical evidence for which 
has been reported by \citet{Kar00}. It is natural
to expect that the probabilities $p_i$, $i=1,...,m$ appear in the generalized
form of
$p_i=(\omega_i\varepsilon_i^{\alpha})/(\sum_{i=1}^m
\omega_i\varepsilon_i^{\alpha})$, where $\varepsilon_i$ are the partial width
of the species and
$\omega_i$ are phenomenological constants. 

Although in the numerical simulation, based on the baker map
as a model flow,
we only found stable coexistence, we also
 carried out simulations where the
biological process was not
 based on parallel stripes filled out homogeneously
with
 individuals, as assumed in our theory.
 In these cellular automaton-like
simulations
 the replication and competition process is carried out
 on a
uniform
 rectangular grid of lattice size
$\epsilon_0$.
This $\epsilon_0$ can be considered as the smallest distance
between the individuals, or the linear size of a single individual
below which there is hard-core exclusion among them.
Individuals of each species can occupy the center of each grid-cell.
When they are advected by the flow into another grid-cell
during the time $\tau$, they are instantaneously placed to the
center of that grid-cell.
During reproduction, they give birth to new individuals in the
surrounding empty grid-cells, whose centers are within a distance
$\sigma_i$.
If more than one species tries to give birth into the same grid-cell,
only one of them will be able to do so according to one of
the following rules:
\emph{rule~I:}~both species
can win this competition in each cell with equal
probability, or
\emph{rule~II:}~both species
can win this competition in each cell with
probability proportional to the number
of individuals of the same species intending to give birth there, or
\emph{rule~III:}~always the better competitor (with higher $\sigma$)
wins.
Our results show that the coexistence depends on which rule
has been applied. In some cases one of the populations was
competed out, but even in such cases the distribution function
was found to be of the shown form, with an exponent
$\alpha >1$, in full harmony with the theory.
It is worth mentioning that with the same rules on the lattice, in
previous simulations \citep{Kar00} for
the more realistic fluid dynamical case
of a flow around an obstacle we always found coexistence.
This indicates that the boundary layer present around the
obstacle enhances the chances of survival.

Our theory does not describe the effects of diffusion. Besides
the fact that for individuals of small but macroscopic mass and
size, like eg.  phytoplankton, diffusion is not believed
to be important, it can be shown \citep{Tel00}
that weak diffusion in such models only renormalizes
the replication rates. As a consequence, the cut-off
scale below which fractality cannot be observed is somewhat
increased, but the population dynamical equations remain unchanged.

In this theory the location dependence
of the death and replication rates
is not taken into account. Such effects can be studied in
numerical simulations \citep{Kar00,San01} 
and are not expected to change the essence of our findings.

In conclusion, we have shown that a particle-like (microscopic) model
of individuals competing for a single resource around a fractal
outflow curve of a chaotic flow leads, on the level of the total
number of individuals, to dynamical equations with unusual
singular terms. These describe enhanced competition due to
inhomogeneous mixing and a kind of advantage of rarity property.
The appearance of unusual population
dynamical equations can be expected in general
in all cases where
the individual dynamics is not taking place on full
compact regions of the space but are
restricted to  fractal subsets of it.

\section*{Acknowledgements}

Z.T. was supported by the Department of Energy 
under contract W-7405-ENG-36.
Support from the Hungarian Science Foundation (OTKA T032423, F029637)
the US-Hungarian Joint Fund (Project No. 501) and the
MTA/OTKA/NSF Fund (Project No. Int. 526) is acknowledged.
G.K.  and I. Sch were supported by the Bolyai grant.
The authors gratefully acknowledge discussions with C. Grebogi,
E. Ben-Naim, T. Cz\'ar\'an and E. Szathm\'ary.

\end{multicols}


\begin{references}

\bibitem[Aref(1994)]{Are94}
Aref, H. (Ed.) 1994:
Special issue on Chaotic Advection.
\emph{Chaos, Solitons, Fractals} \textbf{4}, No.~6.

\bibitem[Ar{\'\i}stegui \etal(1997)]{Ari97}
Ar{\'\i}stegui, J.; Tett, P.; Hern\'andez-Guerra, A.; Basterretxea,
G.; Montero, M.F.; Wild, K.; Sangr\'a, P.; Hern\'andez-Le\'on, S.;
Cant\'on, M.; Garc\'\i{}a-Braun, J.A.; Pacheco, M. \& Barton, E.D. 1997:
The influence of island-generated eddies on chlorophyll distribution:
a study of mesoscale variation around Gran Canaria.
\emph{Deep-See Research}
\textbf{44}, 71--96.

\bibitem[Bartha \etal(1997)]{Bar97}
Bartha, S.; Cz\'ar\'an, T. \& Scheuring, I. 1997:
Spatiotemporal scales of non-equilibrium community dynamics:
a methodological challenge.
\emph{N. Z. J. of Ecol.}\ \textbf{21}, 199--206.

\bibitem[Boerlijst \&  Hogeweg(1991)]{Boe91}
Boerlijst, M.C. \&  Hogeweg, P. 1991:
\emph{Physica D} \textbf{48}, 17--28.

\bibitem[Chesson(2000)]{Che00}
Chesson, P. 2000:
Mechanism of maintenance of species diversity.
\emph{Annu.\ Rev.\ Ecol.\ Syst.}\ \textbf{31}, 343--366.

\bibitem[Connell(1978)]{Con78}
Connell, J.H. 1978:
Diversity in tropical rain forests and coral reefs.
\emph{Science} \textbf{199}, 1302--1310.

\bibitem[Cz\'ar\'an \& Szathm\'ary(2000)]{Cza00}
Cz\'ar\'an, T. \& Szathm\'ary, E. 2000:
In:
Dieckmann, U.; Law, R. \& Metz, J.A.J. (Eds):
\emph{The
Geometry of Ecological Interactions: Simplifying Spatial Complexity}.
Cambridge Univ.\ Press,
Cambridge.

\bibitem[Eigen(1971)]{Eig71}
Eigen, M. 1971:
\emph{Naturwissenschaften} \textbf{58}, 465--523.

\bibitem[Eigen \&  Schuster(1979)]{Eig79}
Eigen, M. \&  Schuster, P. 1979:
\emph{The Hypercycle}.
Springer, Berlin.

\bibitem[Gaedeke \& Sommer(1986)]{Gae86}
Gaedeke, U. \& Sommer, U. 1986:
The influence of the frequency of periodic disturbances on the maintenance
of phytoplankton diversity.
\emph{Oecologia} \textbf{71}, 98--102.

\bibitem[Gause \& Witt(1935)]{Gau35}
Gause, G.F. \& Witt, A.A. 1935:
Behavior of mixed populations and the problem of natural selections.
\emph{Am.\ Nat.}\ \textbf{69}, 596--609.

\bibitem[Gurney \& Nisbet(1998)]{Gur98}
Gurney, W.S.C. \& Nisbet, R.M. 1998:
\emph{Ecological Dynamics}. 
Oxford Univ.\ Press, Oxford.

\bibitem[Hardin(1960)]{Har60}
Hardin, G. 1960: The competitive exclusion principle.
\emph{Science} \textbf{131}, 1292--1298.

\bibitem[Holm(1992)]{Hol92}
Holm, N.G. 1992: Marine hydrothermal systems and the origin of life.
\emph{Origins Life Evol.\ Biosphere}
\textbf{22}, 5.

\bibitem[Hsu \etal(1988)]{Hsu88}
Hsu, G.H.; Ott, E \& Grebogi, C. 1988:
\emph{Phys.\ Lett.\ A} \textbf{127}, 199.

\bibitem[Huston(1979)]{Hus79}
Huston, M.A. 1979:
General hypothesis of species diversity.
\emph{Am.\ Nat.}\ \textbf{113}  81--101.

\bibitem[Hutchinson(1961)]{Hut61}
Hutchinson, G.E. 1961:
The paradox of the plankton
\emph{Am.\ Nat.}\ \textbf{95}, 137--147.

\bibitem[Jung \& Ziemniak(1992)]{Jun92} 
Jung, C. \& Ziemniak, E. 1992:
Hamiltonian scattering chaos in a hydrodynamical system.
\emph{J. Phys.\ A} \textbf{25}, 3929--3943.



\bibitem[Kantz \& Grassberger(1985)]{Kan85}
Kantz, H. \& Grassberger, P. 1985:
Repellers, semi-attractors, and long-lived chaotic transients.
\emph{Physica D} \textbf{17}, 75.

\bibitem[K\'arolyi \& T\'el(1997)]{Kar97} 
K\'arolyi, G. \& T\'el, T. 1997:
Chaotic tracer scattering and fractal basin boundaries in a
blinking vortex-sink system.
\emph{Physics Reports} \textbf{290}, 125--147.


\bibitem[K\'arolyi \etal(1999)]{Kar99}
K\'arolyi, G.; P\'entek, \'A.; Toroczkai, Z.; T\'el, T. \&
Grebogi, C. 1999:
Chemical or biological activity in open chaotic flows.
\emph{Phys.\ Rev.\ E} \textbf{59} 5468--5481.

\bibitem[K\'arolyi \etal(2000)]{Kar00}
K\'arolyi, Gy.; P\'entek, \'A.;
Scheuring, I.; T\'el, T. \& Toroczkai, Z. 2000: 
Open chaotic flow: the physics
of species coexistence. 
\emph{Proc.\ Natl.\ Acad.\ Sci.\ USA} 
\textbf{97} 13661--13665.

\bibitem[Lamb(1932)]{Lam32}
Lamb, H. 1932:
\emph{Hydrodynamics}.
Cambridge University Press, Cambridge.

\bibitem[Maynard Smith(1983)]{May83}
Maynard Smith, J. 1983:   Models of evolution.
\emph{Proc.\ Roy.\ Soc.\ London B}
\textbf{219}, 315--325.

\bibitem[Maynard Smith \& Szathm\'ary (1995)]{May95}
Maynard Smith, J. \& Szathm\'ary, E. 1995:
\emph{The major transitions in evolution}. 
Freeman, Spektrum, Oxford.


\bibitem[P\'entek \etal(1996)]{Pen96} 
P\'entek, \'A.; T\'el, T. \& Toroczkai, Z. 1996:
Transient chaotic mixing in open hydrodynamical flows.
\emph{Int.\ J. Chaos and Bifurcations} \textbf{6}, 2619--2625.

\bibitem[Poole \etal(1999)]{Poo99}
Poole, A.; Jeffares, D. \& Penny, D. 1999:
Early evolution: prokaryotes, the
new kids on the block. 
\emph{BioEssays} \textbf{21}, 880--889.

\bibitem[Reynolds(1993)]{Rey93}
Reynolds, C.S. 1993: 
Scales of disturbance and their role in plankton ecology.
\emph{Hydrobiologia} \textbf{249}, 157--171.

\bibitem[Reynolds(1998)]{Rey98}
Reynolds, C.S. 1998:
The state of freshwater ecology.
\emph{Freshwater Biology} \textbf{39}, 741--753.

\bibitem[Santoboni \etal(2001)]{San01}
Santoboni, G.; Nishikawa, T.; Toroczkai, Z. \& Grebogi, C. 2001:
Autocatalytic reactions of active particles with phase distribution. 
\emph{Preprint}.

\bibitem[Scheuring(2000)]{Sch00a}
Scheuring, I. 2000:
Avoiding Catch-22 of early evolution by stepwise
increase in copying fidelity. 
\emph{Selection} \textbf{1--3}, 135--145.

\bibitem[Scheuring \etal(2000)]{Sch00b}
Scheuring, I.; K\'arolyi, G.; P\'entek, \'A.; Toroczkai, Z. \& T\'el, T.
2000: 
A Model  for
resolving the plankton paradox: coexistence in open flow. 
\textit{Freshwater Biology}
\textbf{45}, 123--133.

\bibitem[Shariff \etal(1991)]{Sha91} 
Shariff, K.; Pulliam, T.H. \&  Ottino, J.M. 1991:
A dynamical systems analysis of kinematics in the time-periodic wake of
a circular cylinder.
\emph{Lect.\ Appl.\ Math.}\ \textbf{28}, 613--646.

\bibitem[Sommer \etal(1993)]{Som93}
Sommer, U.; Padis\'ak, J.; Reynolds C.S. \& Juh\'asz-Nagy, P. 1993:
Hutchinson's heritage: the diversity disturbance relationship
in phytoplankton.
\emph{Hydrobiologia} \textbf{249}, 1--8.

\bibitem[Sommerer \etal(1996)]{Som96}
Sommerer, J.C.;  Ku, H.-C. \&  Gilreath,  H.E. 1996:
Experimental evidence for chaotic scattering in a fluid wake.
\emph{Phys.\ Rev.\ Lett.}\ \textbf{77}, 5055--5058.	

\bibitem[T\'el(1990)]{Tel90}
T\'el, T. 1990: 
In:
Bai-Lin, H. (Ed.):
\emph{Directions in chaos}, Vol.~3., pp.~149--211.
World Scientific, Singapore.

\bibitem[T\'el \etal(2000)]{Tel00} 
T\'el, T.; K\'arolyi, G.; P\'entek, \'A.; Scheuring, I.;  Toroczkai, Z.;
Grebogi, C. \& Kadtke, J. 2000:
Chaotic advection, diffusion, and reactions in
open flows. 
\emph{Chaos} \textbf{10}, 89--98.

\bibitem[Tilman \& Pacala(1993)]{Til93}
Tilman, D. \& Pacala, S. 1993:
In:
Ricklefs, R.E. \& Schluter, D.:
\emph{Species diversity in ecological communities:
The maintenance of species richness in plant communities}.
pp.~13--25, 
Univ.\ of Chicago Press.

\bibitem[Toroczkai \etal(1998)]{Tor98}
Toroczkai, Z.; K\'arolyi, G.; P\'entek, \'A.; T\'el, T.
\& Grebogi, C. 1998:
Advection of active particles in open chaotic flows.
\emph{Phys.\ Rev.\ Lett.}\ \textbf{80}, 500--503.

\bibitem[Toroczkai \etal(1997)]{Tor97}
Toroczkai, Z.; K\'arolyi, G.; P\'entek, \'A.; T\'el, T.;
Grebogi, C. \& Yorke, J.A. 1997:
Wada dye boundaries in open
hydrodynamical flows.
\emph{Physica A} \textbf{239}, 235--243.

\bibitem[Toroczkai \etal(2001)]{Tor01}
Toroczkai, Z.; K\'arolyi, G.; P\'entek, \'A. \& T\'el, T. 2001:
Autocatalytic reactions in systems with
hyperbolic mixing: Exact results for the
active baker map.
\emph{J. Phys.\ A}, \textbf{34}, 5215 -- 5235.

\bibitem[W\"achtersh\"auser(1994)]{Wac94}
W\"achtersh\"auser, G. 1994: Life in a ligand sphere.
\emph{Proc.\ Natl.\ Acad.\ Sci.\ USA}
\textbf{91}, 4283--4287.

\bibitem[Wilson(1990)]{Wil90} 
Wilson, J.B. 1990:
Mechanisms of species coexistence: twelve explanations for Hutchinson's
`paradox of the plankton': evidence from New Zealand plant communities.
\emph{N. Z. J. of Ecol.}\ \textbf{43}, 17--42.
\end{references}
\end{document}